\documentstyle[12pt]{article}

\begin{document}

\def\beq{\begin{equation}}
\def\enq{\end{equation}}
\def\su{$SU(2)_{\em l} \times SU(2)_h\times U(1)_Y$\,}
\def\uem{$U(1)_{\rm{em}}$\,}
\def\suu{$SU(2)\times U(1)_Y$\,}
\def\ttz{{\mbox {$t$-$t$-$Z$}}}
\def\bbz{{\mbox {$b$-$b$-$Z$}}}
\def\tta{{\mbox {$t$-$t$-$A$}}}
\def\bba{{\mbox {$b$-$b$-$A$}}}
\def\tbw{{\mbox {$t$-$b$-$W$}}}
\def\tltlz{{\mbox {$t_L$-$t_L$-$Z$}}}
\def\blblz{{\mbox {$b_L$-$b_L$-$Z$}}}
\def\brbrz{{\mbox {$b_R$-$b_R$-$Z$}}}
\def\tlblw{{\mbox {$t_L$-$b_L$-$W$}}}
\def\beq{\begin{equation}}
\def\enq{\end{equation}}
\def\ra{\rightarrow}
\def\D0{D\O~}
%
%
\begin{titlepage}
\begin{flushright}
{MSUHEP-60315\\
March, 1996 }
\end{flushright}
\vspace{0.4cm}
\begin{center} 
\large
{\bf {A Model of Strong Flavor Dynamics for the Top Quark} }
\end{center}
\begin{center}
{\bf Ehab Malkawi, Tim Tait, and C.--P. Yuan}
\end{center}
\begin{center}
{Department of Physics and Astronomy \\
Michigan State University \\
East Lansing, MI 48824}
\end{center} 
\vspace{0.4cm}
\raggedbottom 
\setcounter{page}{1}
\relax
\begin{abstract}
We discuss a model in which the third generation fermions undergo
a different $SU(2)$ weak interaction from the first two generation 
fermions. In general, a flavor changing neutral current interaction 
is expected. Constrained by the precision low energy data, the mass 
($M_{Z'}=M_{W'}$) of the heavy gauge bosons is bounded from below to 
be about $1.1$ TeV for $\alpha_s=0.125$ and about $1.3$ TeV for 
$\alpha_s=0.115$, at the $3\sigma$ level. This model favors a 
larger $R_b$ and a smaller $R_c$ as compared with the Standard Model,
but it does not explain the $R_c$ data. If one takes the $R_b$ data 
seriously, then $M_{Z'}$ is bounded, at the $3\sigma$ level, to be 
$462$ GeV $< M_{Z^{\prime}} \cos\phi < 1481\,\, {\rm{GeV}}$,
where $\cos\phi$ is the mixing angle between the two $SU(2)$'s in
the model. Effects predicted for high energy experiments at the 
Tevatron, LEP140, LEP-II, LHC, and future linear colliders
are also discussed.
\end{abstract}
\vspace{2.0cm}
\end{titlepage}
\pagestyle{plain}
\newpage
\section{Introduction}
\indent

With the exception of a few measurements, 
all the data agree with the predictions of the Standard Model (SM).
Among those measurements are $R_b$ and $R_c$ at LEP with 
deviations of about $3.5\sigma$ and $2.5\sigma$, respectively 
\cite{lep},
and the excess of large $E_t$ jets at the Tevatron \cite{cdfdijet}.
If one takes the above measurements seriously, one can
advocate specific types of new physics which tackle these experimental 
concerns, such as the studies done in Ref.~\cite{Zprime}. 

In this paper we are not restricting our motives to just explaining 
the measurements discussed
above, but more generally, we are driven by 
a theoretical observation, namely
the hierarchy of the fermion mass spectrum.
The relatively large mass of the third generation fermions 
may suggest a dynamical behavior different from
that of the first two generations. 
Here, we consider a model~\cite{lima} in which 
the third generation undergoes 
a different
flavor dynamics from the usual weak interactions proposed in the SM.
We assume this 
flavor dynamics to be associated with a new $SU(2)$ gauged symmetry. 
Therefore, a new spectrum of gauge bosons emerges in this model.
No modifications to
QCD interactions are considered here; this case has been discussed 
elsewhere \cite{hill}.
\section{The Model}
\indent

This model is based on the electroweak symmetry group \su, where  
the third generation of fermions, (the 
top and bottom quarks; $t$ and $b$, 
the tau lepton and its neutrino;
$\tau$ and $\nu_\tau$) can experience a strong flavor interaction
instead of the weak interaction advocated by the SM. In that
case, the
first two generations only feel the weak 
interactions supposedly
equivalent to the SM case. The quantum numbers of the fermions are\\
For the first two generations,\\
\hspace{.5cm} Left-handed quarks\,: $(2,1)_{1/3}$ \, , \hspace{1cm} 
Left-handed leptons\,: $(2,1)_{-1}$ \, . \\
For the third generation \\
\hspace{.5cm} Left-handed quarks\,: $(1,2)_{1/3}$ \, , \hspace{1cm}
Left-handed leptons\,: $(1,2)_{-1}$ \, . \\ 
For all the right-handed fermions we have\\
\hspace{.5cm} Right-handed quarks and leptons\,: $(1,1)_Q$  \, , \\
where $Qe$ is the electric charge of the right-handed fermions.

The prescribed model is similar to the Ununified Standard Model 
\cite{ununi} in the gauge sector. The 
difference between the two models lies in the fermionic quantum numbers 
under the gauge group. Following the same 
notation in Ref.~\cite{ununi} the covariant derivative may be written as
\beq
\partial^\mu + ig_{\em l} T^a_l W^{\mu}_{la} + ig_h T^a_h W^{\mu}_{ha}+
         ig^{\prime} Y B^{\mu},
\enq
where $T^a_{\em l}$ and $T^a_h$, $a=1$ to 3 
are the $SU(2)$ generators and 
$Y$ is the hypercharge generator. The gauge couplings may be 
written as
\beq
g_l=\frac{e}{\sin \theta \cos \phi}\, , \hspace{1cm} 
g_h=\frac{e}{\sin \theta \sin \phi}\, , \hspace{1cm} 
g^{\prime}=\frac{e}{\cos \theta} \, ,
\enq 
where $\theta$ plays the role of the usual weak mixing angle and 
$\phi$ is a new parameter in this model.

The symmetry  breaking  of the gauge group 
into the electromagnetic group 
\uem is a two-stage mechanism: first,  
\su breaks down into $SU(2)_{{\em l}+h} \times U(1)_Y$ at some large mass scale;
second, $SU(2)_{{\em l}+h}\times U(1)_Y$ breaks down into \uem at
a scale of the order of the SM electroweak symmetry-breaking scale.
This breakdown can be accomplished by 
introducing  two scalar matrix fields: 
{\mbox {$\Sigma=\sigma +i\pi^a\tau^a$}}  and $\Phi$,
with the transformations $\Sigma\sim (2,2)_0$ and $\Phi \sim (2,1)_1$.
The $\tau^a$'s are the 
Pauli 
matrices which satisfy  ${\rm{Tr}}(\tau^a \tau^b)=1/2\delta_{ab}$.
When the $\Sigma$ field acquires a vacuum expectation value (v.e.v.) $u$, 
the symmetry \su is broken into the group
 $SU(2)_{{\em l}+h}\times U(1)_Y$, and the
symmetry-breaking scale is set by the v.e.v $u$.
Then, at a lower energy scale $v$, the 
scalar $\Phi$ acquires a v.e.v $v$
and the symmetry is finally reduced to $U(1)_{\rm{em}}$.
The scalar fields, except for the two remaining neutral Higgs particles,
become the longitudinal components of the physical gauge bosons. 
  
To obtain the gauge boson mass eigenstates, we first write 
the gauge bosons in the following basis:
\beq
{W_1^\pm}_\mu =\cos\phi {W_{\em l}^\pm}_\mu +\sin\phi {W_h^\pm}_\mu\, , 
\hspace{1cm} 
{W_2^\pm}_\mu =-\sin \phi {W_{\em l}^\pm}_\mu +\cos\phi {W_h^\pm}_\mu \, ,
\enq
\beq
{Z_1}_\mu =\cos\theta (\cos\phi {W_{\em l}^3}_\mu +\sin\phi {W_h^3}_\mu)
                -\sin\theta B_\mu \, ,
\enq
\beq
{Z_2}_\mu =-\sin \phi {W_{\em l}^3}_\mu+\cos\phi {W_h^3}_\mu \, ,
\enq
\beq
A_\mu =\sin\theta(\cos \phi {W_{\em l}^3}_\mu +\sin \phi {W_h^3}_\mu)
                +\cos\theta B_\mu\, ,
\enq
where ${W_{\em l}^\pm}_\mu= 
({W_{\em l}^1}_\mu \mp i{W_{\em l}^2}_\mu)/\sqrt{2}$, and similarly 
for ${W_h^\pm}_\mu$. 
The gauge field $A_\mu$ is massless, corresponding to the
physical photon field. 

So far, $g_h$, $g_{\em l}$, and $x$ are free parameters. In this paper
we mainly concentrate on the case where $g_h > g_{\em l}$ (equivalently 
$\tan\phi< 1$) but with $g_h^2 \leq 4\pi$ (which implies
$\sin^2\phi \geq g^2/(4 \pi) \sim 1/30$) so that  
the perturbation theory is valid.
Similarly, for $g_h < g_{\em l}$, we require
$\sin^2\phi \leq 0.96$.
Furthermore, we focus on the region where $x\gg 1$,
though another region of 
interest could be $x\sim 1$ ($u\sim v$), but in this later case
the one-loop level contributions due to 
the heavy gauge bosons  should also be 
included because they are of the 
same order as the SM one-loop contributions.

To order $1/x$ the eigenstates of the light gauge bosons are 
\beq
W_\mu^\pm={W_1^\pm}_\mu +\frac{\sin^3 \phi \cos \phi}{x} 
                             {W_2^\pm}_\mu \, ,
\hspace{.5cm}
Z_\mu={Z_1}_\mu +\frac{\sin^3 \phi \cos \phi}{x\cos \theta} 
                             {Z_2}_{\mu} \, .
\enq
While for the heavy gauge bosons we find
\beq
{W^\prime}_\mu^\pm= -\frac{\sin^3\phi\cos\phi}{x} {W_1^{\pm}}_{\mu} 
                           +{W_2^{\pm}}_{\mu} \, ,
\hspace{.5cm}
Z_\mu^{\prime}= -\frac{\sin^3\phi\cos\phi}{x\cos\theta} {Z_1}_{\mu} 
                           +{Z_2}_\mu \, .
\enq
To the same order, the gauge boson masses are
\beq
M^2_{W^\pm}=M_0^2(1-\frac{\sin^4 \phi}{x} ) \, ,
\hspace{1cm}
M_Z^2=\frac{M_0^2}{\cos^2\theta}(1-\frac{\sin^4 \phi}{x}) \, ,
\enq
\beq
M^2_{{W^\prime}^\pm}=M^2_{Z^\prime}=M_0^2\left ( 
 \frac{x}{\sin^2 \phi \cos^2 \phi} +
       \frac{\sin^2 \phi}{\cos^2 \phi}\right ) \, ,
\enq
where, $M_0=e v/(2\sin\theta)$.
Notice that the heavy gauge bosons are degenerate up 
to this order, i.e. $M_{Z^\prime}=M_{{W^\prime}^\pm}$.
The left-handed fermion couplings to the 
light gauge bosons may be written
as,
\beq
\frac{e}{\sin\theta}\left(T^{\pm}_h+T^{\pm}_{\em l} 
+\frac{\sin^2\phi}{x}
            \left(\cos^2\phi T^{\pm}_h -\sin^2\phi T^{\pm}_{\em l}
\right)\right),
\enq
\beq
\frac{e}{\sin\theta \cos\theta}\left(T^{3}_h+T^{3}_{\em l}-
Q\sin^2\theta +
            \frac{\sin^2\phi}{x}
            \left(\cos^2\phi T^{3}_h -\sin^2\phi T^{3}_{\em l}
\right)\right).
\enq
While the left-handed fermion couplings to the heavy gauge bosons are
\beq
\frac{e}{\sin\theta}\left(\frac{\cos\phi}{\sin\phi} T^{\pm}_h -
                          \frac{\sin\phi}{\cos\phi} T^{\pm}_{\em l} -
                          \frac{\sin^3\phi \cos\phi}{x} \left(
                          T^{\pm}_h + T^{\pm}_{\em l}\right)\right),
\enq
\beq
\frac{e}{\sin\theta}\left(\frac{\cos\phi}{\sin\phi} T^{3}_h -
                          \frac{\sin\phi}{\cos\phi} T^{3}_{\em l} -
                          \frac{\sin^3\phi \cos\phi}{x\cos^2\theta}
\left(
                        T^{3}_h + T^{3}_{\em l} -Q\sin^2\theta \right)
\right).
\enq
The right-handed fermion couplings to the neutral gauge bosons $Z$ and
$Z^\prime$ are, respectively,
\beq
\frac{e}{\sin\theta\cos\theta}\left(-Q\sin^2\theta \right),
\enq
\beq
 \frac{e}{\sin\theta}\left( -Q\sin^2\theta  
   \frac{\sin^3\phi\cos\phi}{x\cos^2\theta}\right).
\enq
The fermion couplings to the photon 
are the usual electromagnetic couplings.
As we see, if $g_h > g_{\em l}$ the heavy gauge bosons 
would couple strongly to the third generation and weakly to the 
first two generations, and vice versa.

The first and second
generations acquire their masses through the Yukawa interactions 
to the $\Phi$ field just as in the SM.
For the third generation we cannot generate their masses through
the usual Yukawa terms (dimension four operators), as it is not allowed 
by gauge invariance. It is only
through higher dimension operators that we 
can generate these fermion masses. 
(Another possible model is to introduce an additional Higgs doublet
as done in Ref.~\cite{lima}.)
This may be attributed to the strong flavor 
dynamics which may be evident at adequately high energies where the 
interactions become strong.
However, we do not offer an
explicit scenario in this paper for such a picture.

Once we generate 
the fermion mass matrices, 
we can  diagonalize them by using bilinear unitary transformations,
and then obtain the physical masses.
Since the third family interacts differently from the first two 
families, we expect in general that Flavor 
Changing Neutral Currents (FCNC)
will occur at tree level.
For the lepton sector we introduce the unitary 
matrices $L_e$, and $R_e$ with the transformations, 
\beq
e_L^i \ra L_e^{ij}e_L^j \, , \hspace{1cm} e_R^i \ra R_e^{ij}e_R^j \, .
\enq
Hence, the physical mass matrix is given by
\beq
M_e^{\rm{diag.}}=L_e^{\dagger}M_e R_e \, .
\enq
We see that new features are manifest in this model, e.g. lepton mixing.
This is an exciting possibility.
Even though the neutrinos are massless,
they can still mix in this model due to the different interactions of
different family neutrinos.
This might be connected 
to the solar neutrino problem.
For the quark sector, we find that  the 
measurements of $R_b$ and $R_c$ from 
the SM favor no mixing in the neutral
sector. Thus we consider FCNCs only in the lepton sector,
but not in the quark sector.

\section{ Low Energy Constraints }
\indent

To test this model by low energy data, 
it is convenient to consider the form of the four-fermion
current-current interactions at zero momentum transfer. 
The four-fermion charged-current weak interactions are
\cite{ununi}
\beq
\frac{2}{v^2}(j_l^\pm+j_h^\pm)^2 + \frac{2}{u^2}{j_h^\pm}^2 \, ,
\label{cc}
\enq
and the neutral current four-fermion interactions are
\beq
\frac{2}{v^2}(j^3_l+j^3_h-\sin^2\theta j_{\rm{em}})^2 + 
\frac{2}{u^2}(j^3_h-\sin^2\phi \sin^2\theta 
j_{\rm{em}})^2 \, ,
\label{nc}
\enq      
where $j_{l,h}^\pm$ are the left-handed charged currents corresponding
to the first two generations and the third generation, respectively.  
Similarly, $j^3_{l,h}$ refers to the left-handed $T^3$ currents, 
while $j_{\rm{em}}$ represents the full electromagnetic 
current of the three 
families.
We conclude that if there is no mixing in the lepton families 
then all leptonic decays are identical 
to the SM, e.g. the $\tau$ lifetime 
can not furnish any new information about this model. 
However, it is more general to allow mixing in the leptonic families, so
we will investigate this possibility more carefully.
Because of the almost vanishing branching ratio of the decay 
$\Gamma{(\mu^{-} \ra e^{-}  e^{+}  e^{-})}$, 
$BR\leq 10^{-12}$ \cite{data}, we will only allow mixing
of $\mu$ and $\tau$ and their neutrinos,
which
depends on one free parameter, $\sin^2\beta$.
The constraints on $\sin^2\beta$ come from: 
the ALEPH measurement
(in terms of the effective couplings ratio $g_\tau/g_\mu$, cf. Table 1)
 of the branching fraction for $\tau$ decay into 
$\mu$ and the determination of the $\tau$ lifetime \cite{aleph}, 
the lepton number violation decay of $\tau\ra \mu\mu\mu$, 
with a branching ratio
 $BR < 4.3\times 10^{-6}$ (at 90\% C.L.) \cite{data}, 
and
the FCNC search at 
LEP with $BR(Z\ra \mu^\pm \tau^\mp) < 1.7\times 10^{-5}$
(at 95\% C.L.) 
\cite{opal}. 
All other fermionic processes at zero momentum transfer,
such as the $\mu$ decay, $K$-$\overline{K}$ mixing, and
$B$-$\overline{B}$ mixing, are identical to the SM
predictions.

In this model, the low energy predictions depend on the values of
$1/x$, $\sin^2\phi$, and $\sin^2\beta$ in addition 
to the measured values of 
$\alpha_{\rm{em}}(M_Z)$, $G_F$, and $M_Z$.
Using the most recent 
LEP measurements \cite{lep} (the total width of the 
$Z$ boson, $R_e$, $R_\mu$, $R_\tau$, 
the vector $g_V$ and axial-vector $g_A$ 
couplings of 
$e$, the ratios $g_V(\mu,\tau)/g_V(e)$, $g_A(\mu,\tau)/g_A(e)$, 
the lepton forward-backward asymmetries, the
$\tau$ and $e$ polarization asymmetry, 
the hadronic pole cross section 
$\sigma_h^0$, and the ALEPH measurement of $g_\tau/g_\mu$ \cite{aleph})
combined with the FCNC measurements 
of $\tau^{-}\ra \mu^{-}\mu^{-}\mu^{+}$
and $Z\ra \mu^{-}\tau^{+}$
we determine the
allowed values of $\sin^2\phi$, $\sin^2\beta$, and $M_{Z^{\prime}}$.
We do not include the controversial 
observables $R_b$ and $R_c$ as a part 
of our fit. Instead, we treat them as a prediction and discuss
later whether our model is able to explain the anomaly in these 
measurements. The experimental values of the electroweak observables 
\cite{lep} and their SM prediction 
\cite{Hagiwara} are given in Table 1. 

We calculate the changes in the relevant physical 
observables relative to their SM values to leading order in $1/x$, i.e.
\beq
O=O^{\rm{SM}}\left(1+\delta O\right ) \, ,
\enq                  
where $O^{\rm{SM}}$ is the SM value for the observable $O$ including the
one-loop SM correction, and $\delta O$ 
represents the new physics effect to leading order in $1/x$.
We list the calculated observables as follows,
\beq
\Gamma_Z = \Gamma_Z^{\rm{SM}}\left( 1+\frac{1}{x}\left[-0.896\sin^4\phi 
               + 0.588\sin^2\phi\right]\right)\, ,
\enq
\beq
R_e =R_e^{\rm{SM}}\left(1 + \frac{1}{x}\left[0.0794\sin^4\phi + 
           0.549\sin^2\phi\right]\right)\, ,
\enq
\beq
R_\mu =R_\mu^{\rm{SM}}\left(1 + \frac{1}{x}\left[0.0794\sin^4\phi +
           0.549\sin^2\phi -2.139\sin^2\beta 
\sin^2\phi\right]\right)\, ,
\enq
\beq
R_\tau =R_\tau^{\rm{SM}}\left(1 + \frac{1}{x}\left[0.0794\sin^4\phi +
           0.549\sin^2\phi -2.139\cos^2\beta 
\sin^2\phi\right]\right)\, ,
\enq
\beq
A_{FB}^e={(A^e_{FB}})^{\rm{SM}}\left(1+
\frac{1}{x}\left[10.44\sin^4\phi\right]
          \right)\, ,
\enq
\beq
A_{FB}^\mu={(A^\mu_{FB})}^{\rm{SM}}
\left(1+\frac{1}{x}\left[10.44\sin^4\phi +
               12.14\sin^2\beta\sin^2\phi\right]\right)\, ,
\enq
\beq
A_{FB}^\tau={(A^\tau_{FB})}^{\rm{SM}}\left(1+\frac{1}{x}\left[10.44
            \sin^4\phi +12.14\cos^2\beta\sin^2\phi\right]\right)\, ,
\enq
\beq
A_e =A_e^{\rm{SM}}\left(1+\frac{1}{x}\left[ 
5.22\sin^4\phi\right]\right)\, ,
\enq
\beq
A_\tau =A_\tau^{\rm{SM}}\left(1+\frac{1}{x}\left[ 5.22\sin^4\phi
                      +12.14\cos^2\beta\sin^2\phi\right]\right)\, ,
\enq
\beq
\sigma_h^0 ={(\sigma_h^0)}^{\rm{SM}} 
\left(1+\frac{1}{x}\left[ -0.01 \sin^4\phi
            -0.628\sin^2\phi\right]\right)\, ,
\enq
\beq
M_W = M_W^{\rm{SM}}\left(1+\frac{1}{x}
\left[1+0.215\sin^4\phi\right]\right)\, ,
\enq
\beq
\frac{g_\tau}{g_\mu}= {(\frac{g_\tau}{g_\mu})}^{\rm{SM}}\left(1+
                   \frac{1}{x}\left[0.50\sin^2\beta
                      \cos^2\beta\right]\right)\, ,
\enq
\beq
{\rm{BR}}(\tau^{-}\ra \mu^{-}\mu^{-}\mu^{+})=
0.045\frac{\sin^2\beta\cos^2\beta}
 {x^2} {\left( \sin^2\beta -4\sin^2\theta\sin^2\phi\right )}^2\, ,
\enq
\beq
\Gamma{(Z\ra \mu^{-}\tau^{+})}= 0.167\,{\rm {GeV}}
             \left ( \frac{\sin^2\phi\sin\beta\cos\beta}{x}
\right)^{2}\, .
\enq

In Figure 1 we show the fit result, at the $3\sigma$ level, of the 
$Z^{\prime}$ mass as a function of $\sin^2\phi$, for 
$\alpha_s=0.125$ and for three 
values of the mixing parameter $\sin^2\beta=0$ (dashed line),
0.5 (dot-dashed line) and 1 
(solid line). In the case of $\sin^2\beta=0$ we find a lower bound on
$M_{Z^\prime}$ of approximately 1.1 TeV. For $\sin^2\beta=1$, 
$M_{Z^\prime}$ is approximately 1.4 TeV. 
For $\sin^2\beta=0.5$, $M_{Z'}$ is required to be larger for smaller
$\sin^2\phi$ ($<0.1$) due to the 
strong constraint from the lepton number 
violating process $\tau \rightarrow \mu \mu \mu$.
As shown in Figure 1, as 
$\sin^2\phi$ increases the lower bound on $M_{Z^\prime}$ increases, and
increase in $M_{Z^\prime}$ is slow for $\sin^2\phi<0.5$ and fast in the
other case. 
This indicates that a relatively light $Z'$ prefers strong 
interactions with the third family fermions.
If we consider a $2\sigma$ fit, then 
the lower bound on $M_{Z^\prime}$ is
about 1.4 TeV for $\sin^2\beta=0$ and 1.8 TeV for $\sin^2\beta=1$. 
For $\alpha_s=0.115$ we find that $M_{Z^\prime}\geq 1.3$ TeV for
$\sin^2\beta =0$ and $M_{Z^\prime}\geq 2.1$ TeV for
$\sin^2\beta =1$ at the 3$\sigma$ level.
   
The LEP measured quantities $R_b=\Gamma_b/\Gamma_h$
and $R_c=\Gamma_c/\Gamma_h$ are not consistent with the SM prediction. 
One possibility to explain the anomaly in 
these quantities is to consider  
new physics  which can affect the $b$
and $c$ quarks' couplings to the $Z$ boson. The question 
now is whether our model is able to give 
any insight regarding these measurements. 
The observed value  
$R_b^{\rm{exp}}=0.2219\pm 0.0017$ \cite{lep} is higher than 
the SM value $R_b^{\rm{SM}}=0.2157$ \cite{Hagiwara} by 
about $3.5\sigma$. 
On the other hand, $R_c^{\rm {exp}}=0.1543\pm 0.0074$ is 
smaller than the SM value $R_c^{\rm{SM}}=0.1721$ by about $2.5\sigma$. 
With the
allowed region of our parameter space being determined, we investigate 
which part of the allowed space is able to explain the anomaly in $R_b$.
Because the measured value of $R_b$ is different from the SM value by
more than $3\sigma$, we expect to be able to constrain the smallest
and largest $Z^{\prime}$ mass by requiring that the new physics effect 
shifts the theoretical value of $R_b$ to be within the $3\sigma$ range
of the measured value.
In our model $R_b$ is given by
\beq
R_b=R_b^{\rm{SM}}\left(1 +\frac{1}{x}\left[-0.0149\sin^4\phi + 
             1.739\sin^2\phi\right]\right)\, .
\enq
Thus, the $Z^{\prime}$ mass can be constrained to be
\beq
 462 \,\,{\rm GeV}  <  M_{Z^{\prime}} \cos\phi < 1481\,\, {\rm{GeV}}\, .
\enq 
Therefore, 
if we assume the anomaly in 
$R_b$ is mainly due to this type of new physics,
then there is an upper bound 
on $M_{Z^\prime}$ which depends on the gauge
coupling (equivalently $\sin\phi$).
For example, for $\sin^2 \phi=0.04$, the upper bound 
(which is  independent of $\sin^2\beta$)
on $M_{Z'}$ obtained from $R_b$ is $\sim 1.5$ TeV.

For $R_c$ we find that the new modification to the SM model shifts $R_c$
in the correct direction, i.e. it decreases the theoretical value as 
desired. However, the amount of shift is too small to account for
its anomaly, e.g. with the lower 
bound on the heavy mass 
coming from 1.1 TeV, we find that the theoretical 
value of 
$R_c$ is still outside the $2\sigma$ range of the measured value.

{}From these results we conclude that this model 
can account for the deviation in $R_b$ from the SM at 
the $3\sigma$ level. Even 
though $R_c$ is shifted in the needed direction, the predicted value
is still outside the $2\sigma$ range of the data.
Therefore, we cannot explain the anomaly in $R_c$ 
entirely based on the proposed model. 
Furthermore, $A_{\rm {LR}}$ in this model is identical to $A_e$. Thus,
this model cannot explain the discrepancy between the
the SLD measurement $A_{\rm {LR}}=0.1551\pm 0.0040$ 
and the LEP measurement
$A_e$ \cite{Hagiwara}. 

\section{ High Energy Experiments }
\indent
 
LEP was operating at the $Z$-pole with large production rates, 
it is therefore unlikely to better test this model 
at other high energy colliders at the scale of $M_Z$.
We have checked that the
measurements of $W^\pm$ and $Z$ properties
at the Tevatron by CDF and \D0 groups \cite{cdfd0} 
do not further constrain
the allowed parameters in 
Figure 1.
To study the possible effects due to the heavy 
$W'$ and $Z'$ bosons, we shall concentrate on physics at
energy scales larger than $M_Z$. 
In this study, the interference effects from 
$\gamma$, $Z$ and $Z'$ in neutral channels and the 
interference of $W$ and $W'$ in charged channels are all included.
To simplify our discussion, we shall consider two sets of parameters
for ($x$, $\sin^2\phi$, $\sin^2\beta$)~: (7,0.04,0) and
(20.6,0.14,0.5) which correspond to ($M_{Z'}$,$\Gamma_{Z'}$)
equal to (1083,291)\,GeV and (1050,76)\,GeV, respectively.
Our conclusions, however,
will not significantly depend on the details
of the parameters chosen from Figure 1.

At the Tevatron, it is possible to reach the high energy region
where the $W'$ or $Z'$ effects can be important. CDF has reported 
the result of searching for new gauge bosons by measuring the 
number of excess di-lepton events with large 
transverse mass~\cite{chsearch} or  invariant mass~\cite{nsearch}.
We find that those results do not further constrain the parameters 
shown in Figure 1.
For the Tevatron with Main Injector
(a ${\rm {\bar p}}{\rm p}$  collider at 
$\sqrt{S}=2$\,TeV with a $2\,{\rm fb}^{-1}$ luminosity),
the excess in the $e^-e^+$ or $e^+ \nu_e$ rates from this model 
is generally not big enough to be easily observed.
Since the third family leptons can strongly 
couple to the new gauge bosons,  the rate of $\tau$ lepton  
production can in principle be quite 
different from that of $e$ or $\mu$. 
Furthermore, if $\sin \beta$ is not zero, the production rates of 
${\rm {\bar p}}{\rm p} \rightarrow W,{W'} \rightarrow \ell  
{}\nu_{\ell}$
or
${\rm {\bar p}}{\rm p} \rightarrow \gamma,Z,Z' \rightarrow \ell  
\bar{\ell}$
will be different for $\ell=e \,{\rm and}\, \mu$.
However, even with the maximal mixing between $\tau$ and $\mu$
(i.e. $\sin \beta=1$) this difference at the Tevatron can only exceed 
a $3\sigma$ effect for a $10\,{\rm fb}^{-1}$ of integrated  
luminosity.
At the LHC (a ${\rm p}{\rm p}$ collider with $\sqrt{S}=10$\,TeV
and a luminosity of $100\,{\rm fb}^{-1}$), this excess cannot be
mistaken. Furthermore, at the LHC, the excess in the 
production rates of the 
$\ell \nu_\ell$ and the $\ell^+\ell^-$ events can also be  
individually tested.
Thus, it is much easier to either find such 
new effects or constrain parameters of the model
at the LHC than at the Tevatron.
We note that this conclusion holds for either a small or large 
$\sin ^2 \phi$. Although with a large $\sin^2 \phi$, the new physics 
effects to light family fermions will
be large, because of the large $W'$ and $Z'$ masses, the net 
effect of the new physics to the production of di-lepton pairs 
does not significantly depend on  $\sin^2\phi$.

Another signature of the model is
an excess in the top quark production,
however, this excess cannot be observed at the Tevatron because of 
large background from the QCD processes 
$q \bar{q}, gg \rightarrow t \bar{t}$.
At the LHC, the excess in the $t \bar{t}$ pair productions 
can easily be seen in the invariant mass
distributions.
The extra gauge bosons can also produce an excess of di-jet events
in the large invariant mass region, but  the parameter space remaining
after imposing low energy constraints does not allow a big enough effect
to explain the results reported by CDF~\cite{cdfdijet}.

Another possible interesting signature is the production of 
$\mu^\pm \tau^\mp$ pairs, which is unconstrained by current LEP data.
At the Tevatron for ($x,\sin^2 \phi,\sin^2\beta$) equal to  
(20.6,0.14,0.5),
the most favorable scenario for observing this signal,
we find a total of about 20 events for $2\,{\rm fb}^{-1}$ of  
integrated
Luminosity, assuming no cuts are imposed.   It is interesting to notice
that this implies that the upgraded Tevatron can provide a
better constraint on this FCNC type of event than LEP can.
At the LHC, the cross section is big,
about 170 fb for this choice  
of parameters.

At high energy electron colliders, the detection of the
above new signatures becomes much easier as long as there are
enough of them produced in the collisions.
In this model, neither LEP140 or LEP-II can see them,
so we shall concentrate on the 
future high energy Linear Collider (LC)~\cite{linearc}.
Consider the proposed $e^{+} e^{-}$ LC at center of mass (CM)
energy $\sqrt{s}=500$ GeV with an integrated luminosity of 
50 ${\rm{fb}}^{-1}$.
For $m_t=175$ GeV the SM production rate 
$\sigma^{SM}_{(e^{+} e^{-} \ra t\overline{t})}$
is 558 fb. Thus a large number of $t$-$\overline{t}$  pairs is  
expected at the NLC.
Considering the set of
parameters ($x$,$\sin^2\phi$,$\sin^2\beta$) = (7.0,0.04,0.0),      
we find that $\sigma_(e^{+} e^{-} \ra t\overline{t})
=709$ fb, i.e. there is about
27\% increase in the total production rate compared to the SM. 
At the LC it is expected
to measure the $t$-$\overline{t}$ cross section, for 
$\ell +\,{\rm jets}$ decay modes,
to within a few percent. 
With the assumption that the expected measurement is within 3  
standard 
deviation from the SM, we can constrain the parameters 
to those which produce $M_{Z'} \geq$ 2.3 TeV.
We note that the same constraints hold for 
different choices of $\sin^2\phi$ and $x$ but with almost the 
same ratio $\sin^2\phi/x$, especially for small
$\sin\phi$, since in the cross section the two parameters enter as 
a ratio.
Because only the left-handed couplings of the top quark
are significantly modified in this model, measuring 
the angular distribution of $t$ in the $t$-$\overline{t}$
CM frame, or its production rate from a polarized $e^\pm$ beam,
can further improve these bounds if no new signal is
found.

Although the $e^{+} e^{-}$ LC is suitable to probe the model under  
study, 
we notice that the $\mu^{+} \mu^{-}$ collider is also interesting
because of the possible mixing between $\mu$ and $\tau$ leptons.
For small mixing the  $e^{+} e^{-}$ and 
the $\mu^{+} \mu^{-}$ colliders lead to similar 
production rates as expected. 
For large $\sin \beta$ the total production rate of 
$\sigma_(\mu^{+} \mu^{-} \ra t\overline{t})$ becomes smaller than the 
SM rate which shows the opposite effect to the production of   
the $e^+e^- \rightarrow t \bar t$ events predicted by this model.
For the same reason,
it is easy to observe the difference in the production rates
of $e^-e^+$ and $\mu^- \mu^+$ (or $\tau^+ \tau^-$)
pairs at the LC. 
Furthermore, at the LC, if the 
FCNC event $e^-e^+ \rightarrow \mu^\pm \tau^\mp$  
occurs,
it can be unmistakably identified. For a 500\,GeV LC with 
a $50\,{\rm fb}^{-1}$ luminosity, we expect an order of  300
such events to be
observed for ($x,\sin^2 \phi,\sin^2\beta$) equal to (20.6,0.14,0.5).
Figure 2 shows the FCNC event numbers at the LC for a few
choices of parameters, assuming no cuts are imposed.

In summary, we find that due to the strong constraints to this model
implied from low energy data (including $Z$-pole data)
it is not easy to find events with new signatures predicted for 
Tevatron or LEP-II. However, at the LHC and the LC, it becomes 
easy to detect deviations from the SM in the productions of 
the third family or second family (in case of 
large mixing between $\tau$ and $\mu$ lepton) fermions.
We have also checked the possible excess in the $W^+W^-$ or the  
$W^\pm Z$
productions at future high energy colliders. 
It turns out that the branching ratios for $Z'$ or $W'$ to the pure  
gauge
boson modes are always small, so the gauge boson pair productions
are not good channels for testing this model.

In the process of preparing for this paper, we noticed that another
similar work was done in Ref.~\cite{muller}. Our conclusions on the 
allowed parameters of the model and the predictions on the event  
yields
for electron or hadron colliders are different from theirs.

\section{ Acknowledgements }
\indent

We thank G.L. Kane, F. Larios, and C. Schmidt for helpful discussions.
This work was supported in part by the NSF under grant no. PHY-9309902.

\newpage

\section*{Table Captions}
Table. 1.\\
Experimental [1,8] and predicted values of electroweak
observables for the SM [10] and the proposed model
(with different choices of parameters)
for $\alpha_s=0.125$ with $m_t=175$ GeV and $m_H=300$ GeV.\\

\vspace{2.0in}

\section*{Figure Captions}
Fig. 1.\\
The lower bound on the heavy $Z^\prime$ mass 
as a function of $\sin^2\phi$
at the $3\sigma$ level , for $\sin^2\beta=0$ (solid),
$\sin^2\beta=1$ (dashed),
and  $\sin^2\beta=0.5$ (dot-dashed) with
$\alpha_s=0.125$.\\

\noindent
Fig. 2.\\
The number of $\mu^\pm \tau^\mp$ events produced at the LC,
an $e^+ e^-$ collider at $\sqrt{s}=500$ GeV with an integrated luminosity of
50 ${\rm fb}^{-1}$ as a function of $\sin^2\beta$, for two choices of parameters:
($M_{Z'}$, $\sin^2 \phi$, $\Gamma_{Z'}$) = (1700 GeV, 0.1, 174 GeV) and
(1700 GeV, 0.4, 55 GeV).
\\

\newpage

\begin{table}
\caption{Experimental [1,8] and predicted values of electroweak
observables for the SM [10] and the proposed model
(with different choices of parameters)
for $\alpha_s=0.125$ with $m_t=175$ GeV and $m_H=300$ GeV.}

\vspace{0.5in}

\begin{tabular}{|c|c|c|c|c|c|c|}  \hline\hline
Observables&Experimental data&SM
&\multicolumn{4}{|c|} {\mbox {The model}}\\ \hline
     &          &  & a & b & c & d  \\ \hline
$g_V(e)$          &   $-0.0368\pm 0.0017$   
&$-0.0367$ &$-0.0367$&$-0.0367$&$-0.0372$ &
$-0.0371$\\
$g_A(e)$          &   $-0.50115\pm 0.00052$ 
&$-0.5012$ & $-0.5012$&$-0.5012$&$-0.5005$ &
$-0.5006$ \\
$g_V(\mu)/g_V(e)$ &   $1.01\pm 0.14$        
& 1.00   &1.00   &1.05   &1.00    &
1.04   \\
$g_A(\mu)/g_A(e)$ &   $ 1.0000\pm 0.0018$   
& 1.0000 &1.0000 &1.0034 &1.0000  &
1.0030 \\
$g_V(\tau)/g_V(e)$&   $1.008\pm 0.071$      
& 1.000  &1.073  &1.000  &1.047   &
1.000 \\
$g_A(\tau)/g_A(e)$&   $1.0007\pm 0.0020$    
& 1.0000 &1.0055 &1.0000 &1.0036  &
1.0000 \\
$\Gamma_Z$      & $2.4963\pm 0.0032$  
& 2.4978 & 2.5054 & 2.5025 & 2.4967 &
2.4969\\  
$R_e$           & $20.797\pm 0.058 $  
& 20.784 & 20.848 & 20.823 & 20.830 &
20.822\\  
$R_\mu$         & $20.796\pm 0.043 $  
& 20.784 & 20.848 & 20.671 & 20.830 &
20.690\\  
$R_\tau$        & $20.813\pm 0.061 $  
& 20.831 & 20.648 & 20.870 & 20.717 &
20.869\\  
$\sigma_h^0$    & $41.488\pm 0.078 $  
& 41.437 & 41.293 & 41.348 & 41.343 &
41.359\\  
$A_e$           & $0.139\pm  0.0089$  
& 0.1439 & 0.1441 & 0.1440 & 0.1461 &
0.1457\\ 
$A_\tau$        & $0.1418\pm 0.0075$  
& 0.1439 & 0.1537 & 0.1440 & 0.1523 &
0.1457\\  
$A^{FB}_e$      & $0.0157\pm 0.0028$  
& 0.0157 & 0.0157 & 0.0157 & 0.0162 &
0.0161\\ 
$A^{FB}_\mu$    & $0.0163\pm 0.0016$  
& 0.0157 & 0.0157 & 0.0164 & 0.0162 &
0.0167\\
$A^{FB}_\tau$   & $0.0206\pm 0.0023$  
& 0.0157 & 0.0168 & 0.0157 & 0.0169 &
0.0161\\
$g_\tau/g_\mu$  & $0.9943\pm 0.0065$  
& 1.0000 & 1.0000 & 1.0000 & 1.0000 &
1.0000\\ \hline
$R_b$           & $0.2219\pm 0.0017$  
& 0.2157 & 0.2178 & 0.2170 & 0.2170 &
0.2168\\
$R_c$           & $0.1543\pm 0.0074$  
& 0.1721 & 0.1716 & 0.1718 & 0.1718 &
0.1718\\
$M_W$           & $80.26\pm 0.16   $  
& 80.32  & 80.32  & 80.32  &  80.37 &
80.36\\ 
$A_{LR}$        & $0.1551\pm 0.0040$  
& 0.1439 & 0.1441 & 0.1440 & 0.1461 &
0.1457\\ \hline
\end{tabular}
\end{table}
\noindent
{\footnotesize 
a: $\sin^2\beta=0$, $\sin^2\phi=0.04$, $M_{Z^\prime}=1.1$ TeV,
$\Gamma_{Z^\prime}=288$ GeV.}\\
{\footnotesize
b: $\sin^2\beta=1$, $\sin^2\phi=0.04$, $M_{Z^\prime}=1.4$ TeV,
$\Gamma_{Z^\prime}=370$ GeV.}\\
{\footnotesize
c: $\sin^2\beta=0$, $\sin^2\phi=0.80$, $M_{Z^\prime}=3.0$ TeV,
$\Gamma_{Z^\prime}=287$ GeV.}\\
{\footnotesize
d: $\sin^2\beta=1$, $\sin^2\phi=0.80$, $M_{Z^\prime}=3.3$ TeV,
$\Gamma_{Z^\prime}=316$ GeV.}


\newpage

\begin{thebibliography}{99}
\bibitem{lep}
The LEP Collaborations and the LEP Electroweak Working Group,
CERN-PPE/95-172.
\bibitem{cdfdijet}
A. Bhatti, for the CDF and \D0 Collaborations, FNAL-CONF-95/192-E,
presented at 10th Tropical Workshop on 
Proton-Antiproton Collider Physics, Batavia,
IL, 9-13 May 1995.
\bibitem{Zprime}
G. Altarelli, et al., CERN-TH/96-20, January 1996;\\
P. Chiappetta, et al., PM/96-05, hep-ph/9601306.
\bibitem{lima}
X. Li and E. Ma, Phys. Rev. Lett. {\bf 47}, 
1788 (1988); {\it ibid.} {\bf 60},
495 (1988); Phys. Rev. {\bf D46}, 1905 (1992); 
J. Phys. {\bf G19}, 1265 (1993).
\bibitem{hill}
C. T. Hill, Phys. Lett. {\bf B345}, 483 (1995).
\bibitem{ununi}
H. Georgi, E.E. Jenkins, and E.H. Simmmons, Phys. Rev. Lett {\bf 62},
 2789 (1989); {\em ibid.}, Nucl. Phys. {\bf B331}, 541 (1990);\\
R.S. Chivukula, E.H. Simmons and J. Terning, hep-ph/9412309. 
\bibitem{data}
L. Montanet et al., Phys. Rev. {\bf D50}, 1173 (1994) and
1995 off-year partial update for the 
1996 edition (URL:http://pdg.lbl.gov/).
\bibitem{aleph}
The ALEPH Collaboration, CERN-PPE/95-127. August 1995.
\bibitem{opal}
The OPAL Collaboration, CERN-PPE/95-43, April 1995.
\bibitem{Hagiwara}
K. Hagiwara, KEK-TH-461, Dec. (1995).
\bibitem{cdfd0}
S. Abachi, {\it et al.}, Phys. Rev. Lett. {\bf 75}, 1456 (1995) and references therein; \\
F. Abe, {\it et al.}, Phys. Rev. Lett. {\bf 73}, 220 (1994);
and references therein.
\bibitem{chsearch}
CDF Collaboration, F. Abe et al., Phys. Rev. Lett. {\bf 75}, 2900 (1995).
\bibitem{nsearch}
CDF Collaboration, F. Abe et all., Phys. Rev. {\bf D 51}, 949 (1995).
\bibitem{linearc}
{\em $e^+ e^-$ Collisions at 500\, GeV: The Physics Potential,} Hamburg,
Germany,1993, Ed. P.M. Zerwas; and {\em Physics and Experiments with
Linear $e^+ e^-$ Colliders}, Waikoloa, USA, 
1993, Ed. F.A. Harris, et al.;
and references therein.
\bibitem{muller}
D.J. Muller and S. Nandi, hep-ph/9602390, 1996.
\end{thebibliography}
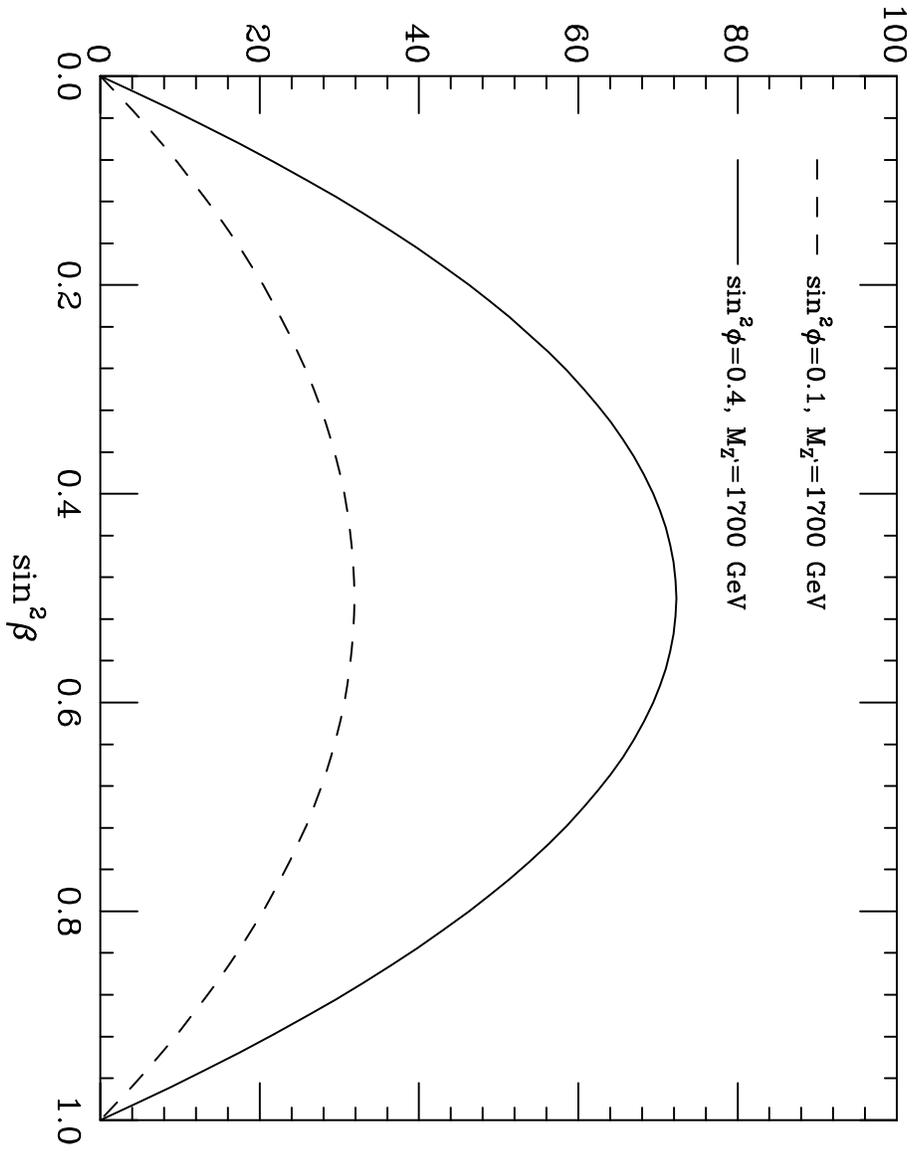
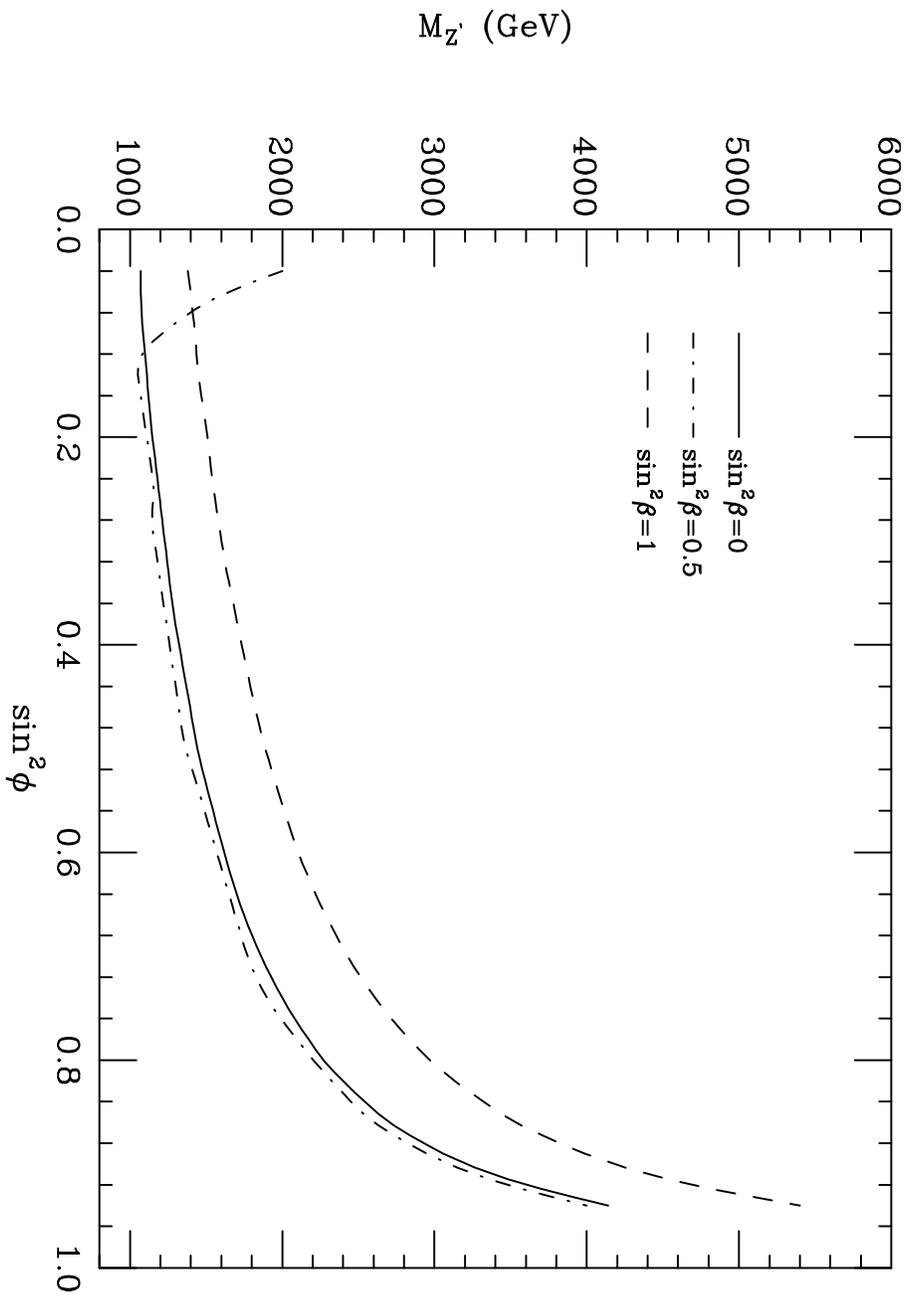
\end{document}